\documentclass[prb,twocolumn,superscriptaddress,showpacs]{revtex4}

\usepackage{pslatex}
\usepackage{amsmath}
\usepackage{amsthm}
\usepackage{amssymb}
\usepackage{amsbsy}
\usepackage{graphicx}
\usepackage{pstricks}
\usepackage{wasysym}
\usepackage{verbatim}
\usepackage[dvips]{epsfig}
\usepackage{pictexwd,dcpic}

\newcommand{\la}{\langle}
\newcommand{\ra}{\rangle}

\begin{document}

\newtheorem{proposition}{Proposition}
\newtheorem{lemma}{Lemma}

% You should use BibTeX and apsrev.bst for references
\bibliographystyle{apsrev}
% Use the \preprint command to place your local institutional report
% number on the title page in preprint mode.
% Multiple \preprint commands are allowed.
%\preprint{}

%Title of paper
\title{Lieb-Schultz-Mattis theorem for quasi-topological systems}
% Optional argument for running titles on pages
%\title[]{}

% repeat the \author .. \affiliation  etc. as needed
% \email, \thanks, \homepage, \altaffiliation all apply to the current
% author. Explanatory text should go in the []'s, actual e-mail
% address or url should go in the {}'s for \email and \homepage.
% Please use the appropriate macro for the type of information

% \affiliation command applies to all authors since the last
% \affiliation command. The \affiliation command should follow the
% other information

\author{Michael Freedman}
\affiliation{Microsoft Research, Station Q, CNSI Building, University of
California, Santa Barbara, CA 93106, USA}
\author{Chetan Nayak}
\affiliation{Microsoft Research, Station Q, CNSI Building, University of
California, Santa Barbara, CA 93106, USA}
\affiliation{Department of Physics, University of California, Santa Barbara,
CA 93106, USA}
\author{Kirill Shtengel}
\email[]{kirill.shtengel@ucr.edu}
\affiliation{Department of Physics, University of California,
Riverside, CA 92521, USA}
\affiliation{California Institute of Technology, Pasadena, CA 91125, USA}
%\homepage[]{Your web page}
%\thanks{}
%\altaffiliation{}
\date{\today}

\begin{abstract}
  In this paper we address the question of the existence of a spectral
  gap in a class of local Hamiltonians. These Hamiltonians have the
  following properties: their ground states are known exactly; all
  equal-time correlation functions of local operators are
  short-ranged; and correlation functions of certain non-local
  operators are critical.  A variational argument shows gaplessness
  with $\omega\propto k^2$ at critical points defined by the absence
  of certain terms in the Hamiltonian, which is remarkable because
  equal-time correlation functions of local operators remain
  \emph{short-ranged}. We call such critical points, in which spatial
  and temporal scaling are radically different,
  \emph{quasi-topological}.  When these terms are present in the
  Hamiltonian, the models are in gapped topological phases which are
  of special interest in the context of topological quantum
  computation.
  \end{abstract}

% insert suggested PACS numbers in braces on next line
%\maketitle must follow title, authors, abstract and \pacs
\maketitle
%%%%%%%%%%%%

\section{Introduction}
Recently, there has been great interest in various unconventional
states of matter, in particular those that might appear in
strongly-interacting 2D systems. These exotic phases may be
characterized by unconventional order parameters. Alternatively, the
phases may be \emph{topological}. In this case there is no local order
parameter, but there may be a \emph{non-local} order parameter, which
is related to the topological properties of the state: exotic braiding
statistics, quantum number fractionalization, and a ground-state (GS)
degeneracy which depends only on the topology of the underlying 2D
manifold \cite{Wen90a}.  The low-energy effective description of these
phases is a topological quantum field theory (TQFT).  A variety of
topological phases are known to exist in the quantum Hall systems
\cite{DasSarma97}. It has been conjectured that such phases also occur
in frustrated magnets, where they may be connected to
superconductivity
\cite{Anderson87,Kivelson87,Read91a,Balents00,Senthil00,Laughlin88a,Chen89}.
Topological phases are also an attractive platform for quantum
computation, where their insensitivity to local perturbations leads to
fault-tolerance \cite{Kitaev97}.  Non-Abelian topological phases are
particularly interesting in this context because, in many of these
phases, the braiding of quasiparticles generates a set of
transformations which is sufficient for universal quantum computation
\cite{Freedman02a}. In this paper, we shall concentrate on microscopic
models which are related to a class of $P-$ and $T-$invariant
non-Abelian topological phases \cite{Freedman04a}.

A set of conditions which places a microscopic model in such a
topological phase can be briefly summarized as follows. We suppose
that the low-energy Hilbert space can be mapped onto that of a quantum
loop gas.  This is the case in a large class of models, including
dimer models, certain spin models, and some interacting hard-core
boson models.  In such a model, basis states are associated to
collections of non-intersecting loops
\cite{Freedman04a,FNS03b,Freedman05b}.  We give some examples in
section \ref{sec:micro-models}.  A Hamiltonian can act on states in
this Hilbert space by doing the following: (i) the loops can be
continuously deformed -- we will call this an isotopy move; (ii) a
small loop can be created or annihilated -- the combined effect of
this move and the isotopy move has been dubbed
``$d$-isotopy''\cite{Freedman03,Freedman04a,Freedman05b}; (iii)
finally, when exactly $k+1$ strands come together in some local
neighborhood, the Hamiltonian can cut them and reconnect the resulting
``loose ends'' pairwise so that the newly-formed loops are still
non-intersecting.  More specifically, in order for this model to be in
a topological phase, the ground state of this Hamiltonian should be a
superposition of all such pictures with the additional requirements
that (i) if two pictures can be continuously deformed into each other,
they enter the GS superposition with the same weight; (ii) the
amplitude of a picture with an additional loop is $d$ times that of a
picture without such loop; (iii) this superposition is annihilated by
the application of the Jones--Wenzl (JW) projector that acts locally
by reconnecting $k+1$ strands (for a detailed description see the
following section).  The main goal of this paper is to investigate the
energy spectrum of a system subject to the first two conditions.

We shall show that a generic local Hamiltonian which enforces
$d$-isotopy for its ground state(s) is necessarily gapless provided
that $ \vert d \vert \leq \sqrt{2}$. (We shall also argue that,
surprisingly, even the addition of the JW projector to the Hamiltonian
will not open a gap for $ \vert d \vert = \sqrt{2}$.)  Such a
Hamiltonian is a sum of projection operators (enforcing on every
plaquette of the lattice both isotopy invariance and the value $d$ for
a contractible loop).  These projection operators do not commute with
each other, but they are compatible with each other in the sense that
they all annihilate the ground state.  (Such Hamiltonians have also
arisen in the context of the quantum Hall effect, where the Haldane
pseudopotentials are projection operators which annihilate the
Laughlin states \cite{Trugman85,Haldane-QHE}, but do not commute so
the excited states are not known exactly, and, under a general name of
``parent Hamiltonians'', in quantum antiferromagnetism
\cite{Majumdar69,Klein82,Affleck87}.) Exact knowledge of the ground
state enables us to construct a variational \emph{ansatz} for the
lowest energy excited state. The strategy is quite similar to the
single-mode approximation (SMA), in which ${\rho_q}|0\rangle$ is the
trial excited state, where $\rho$ is some conserved charge. In the
case of a broken symmetry state, $\rho$ is the charge which generates
the symmetry transformation which is spontaneously broken in the
ground state. Our method generalizes the SMA in an important way: we
do not rely on the existence of any conserved charges and, indeed, the
model need not have any.  Instead, we have the less restrictive
condition that the configuration space of the model should break into
two (or more) sectors whose volume is parametrically larger than the
boundary between them, e.g. by a factor of the system size $L$. If the
Hamiltonian only has matrix elements between nearby points in
configuration space, then (\`{a} la Lieb-Schultz-Mattis \cite{Lieb61})
we can construct a wavefunction which is equal to the ground state
wavefunction except for a relative sign change between the two
sectors. The energy cost of this ``twisted'' excited state would be at
most $\sim 1/L$ and is even smaller in the case of the models which we
consider here.  On the other hand, if the Hamiltonian directly
connects the two putative sectors, then it means that in reality these
are not two distinct sectors, and a sign change in the wavefunction
would have an energetic penalty which does not scale to zero as
$L\rightarrow\infty$.

In this paper, we prove the general result on gaplessness described in
the preceding paragraph. We apply this result to microscopic models
implementing $d$-isotopy and find the remarkable result that they are
gapless with $\omega\propto k^2$ in spite of the absence of any
power-law equal-time correlation functions of local operators.

\section{$d$-isotopy and its local subspaces}
\label{sec:d-isotopy}

In this section we present a more formal definition of $d$-isotopy; it
can be skipped on a first reading. Readers interested in the details
are also referred to Refs.~\onlinecite{Freedman03,Freedman04a,FNS03b}.

\begin{figure}[hbt!]
\includegraphics[width=2.75in]{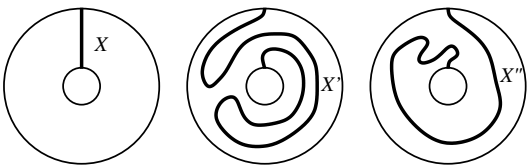}
\caption{Isotopy on
  an annulus: $X\sim X'$ but $X \nsim X''$.}
\label{fig:isotopy}
\end{figure}

Although we will eventually be considering a system on a lattice, it
is useful to begin by defining the Hilbert spaces of interest,
$\overline{V}_d$ and ${V}_d$, in the smooth, lattice-free setting.
Consider a compact surface $Y$ and the set $S$ of all multiloops
\footnote{A multiloop is a collection of non-intersecting loops and
  arcs on a surface. The end points of an arc are required to lie on
  the boundary of the surface, see Ref.~\onlinecite{Freedman04a}.}
$X\subset Y$.  If $\partial Y$ (the boundary of $Y$) is non-empty, we
fix once and for all a finite set $P$ of points on $\partial Y$ with
$X \cap \partial Y =P$.  We assume $Y$ is oriented but $X$ should
\emph{not} be. There is a large vector space $\mathbb{C}^S$, of
complex valued functions on $S$.  We say $X$ and $X'$ are
\emph{isotopic} ($X{\sim}X'$) if one may be gradually deformed into
the other with, of course, the deformation being the identity on
$\partial Y$ (see Fig.~\ref{fig:isotopy}).

%%%%%%%%%%%%%%%%%%%%%%%%%%%%%%%%%%%%%%%%%%%%%%%%%%%%%%%%%%%%
%%%%%%%%%%%%%%%%%%%%%%%%%%%%%%%%%%%%%%%%%%%%%%%%%%%%%%%%%%%%

We may view the isotopy relation as a family of linear constraints on
$\mathbb{C}^S$, namely $\Psi(X)-\Psi(X') =0$ if $X \sim X'$.  The
subspace satisfying these linear constraints is now only of countable
dimension; it consists of those functions which depend only of the
isotopy class $[X]$ of $X$ and may be identified with
$\mathbb{C}^{[S]}$, where $[S]$ is the isotopy classes of multiloops
with fixed boundary conditions.  Note that all isotopes can be
made by composing small locally-supported ones, so the relations we
just imposed are ``local'' in the sense that they can be implemented
with purely local terms in the Hamiltonian.

Let us go further and define an additional local relation which, when
added to isotopy, constitutes the ``$d$-isotopy'' relation.  This
relation is:
\begin{equation}
  \label{eq:d-isotopy}
  d\;\Psi(X)-\Psi(X \cup \bigcirc)=0
\end{equation}

It says that if two multiloops are identical except for the presence
of a small (or, it follows, any contractible) circle, then their
function values differ by a factor of $d$, a fixed positive real
number. In cases of interest to us $1\leq d<2$, so our function is
either neutral to or ``likes'' small circles. We call the subspace
obeying all these constraints the $d$--isotopy space of $Y$ (with
fixed boundary conditions) and write it as $\overline{V}_d \subset
\mathbb{C}^{[S]} \subset\mathbb{C}^S$. The subspace $\overline{V}_d
(T^2)$ is still of countable dimension, or extensively degenerate, on
the torus $T^2$. The vector space $\overline{V}_d$ is clearly the
ground state manifold (GSM) of a local Hamiltonian acting on
$\mathbb{C}^{[S]}$.

It is a remarkable fact \cite{Freedman03,Freedman04a} that it is very
difficult to add any further local relations to $d-$isotopy without
killing the vector space entirely.
For the physically interesting cases $\alpha = \text{e}^{\pi \text{i}/(k+2)}$,
$k=1,2,3, \ldots$, there is such a local
relation and a natural positive definite inner product on
$\overline{V}_d$ (see
refs.~\onlinecite{Freedman03,Freedman04a} for definition).

In these cases the local relations are essentially the Jones-Wenzl
idempotents:\\
$k=1$:
\begin{equation}
  \label{eq:JW1}
  \pspicture[0.4](1.0,1.0)
%   \pspicture[shift=-0.4](1.0,1.0)
  \psframe[linewidth=0.5pt](0,0)(1.0,1.0)
  \psbezier[linewidth=1.0pt](0.333333333333333,0)
  (0.333333333333333,0.5)(0.333333333333333,0.5)(0.333333333333333,1.0)
  \psbezier[linewidth=1.0pt](0.666666666666667,0)(0.666666666666667,0.5)
  (0.666666666666667,0.5)(0.666666666666667,1.0)
  \endpspicture
  \; - \; \frac{1}{d}
  \pspicture[0.4](1.0,1.0)
%  \pspicture[shift=-0.4](1.0,1.0)
  \psframe[linewidth=0.5pt](0,0)(1.0,1.0)
  \psbezier[linewidth=1.0pt](0.333333333333333,0)
  (0.333333333333333,0.333333333333333)
  (0.666666666666667,0.333333333333333)(0.666666666666667,0)
  \psbezier[linewidth=1.0pt](0.666666666666667,1.0)
  (0.666666666666667,0.666666666666667)
  (0.333333333333333,0.666666666666667)(0.333333333333333,1.0)
  \endpspicture
  =0
\end{equation}
$k=2$:
\begin{multline}
  \label{eq:JW2}
  \pspicture[0.4](1.0,1.0)
%  \pspicture[shift=-0.4](1.0,1.0)
  \psframe[linewidth=0.5pt](0,0)(1.0,1.0)
  \psbezier[linewidth=1.0pt](0.25,0)(0.25,0.5)(0.25,0.5)(0.25,1.0)
  \psbezier[linewidth=1.0pt](0.5,0)(0.5,0.5)(0.5,0.5)(0.5,1.0)
  \psbezier[linewidth=1.0pt](0.75,0)(0.75,0.5)(0.75,0.5)(0.75,1.0)
  \endpspicture
  +
  \frac{1}{d^2-1}
  \left(\: 
  \pspicture[0.4](1.0,1.0)
%  \pspicture[shift=-0.4](1.0,1.0)
    \psframe[linewidth=0.5pt](0,0)(1.0,1.0)
    \psbezier[linewidth=1.0pt](0.25,0)(0.25,0.25)(0.5,0.25)(0.5,0)
    \psbezier[linewidth=1.0pt](0.75,0)(0.75,0.5)(0.25,0.5)(0.25,1.0)
    \psbezier[linewidth=1.0pt](0.75,1.0)(0.75,0.75)(0.5,0.75)(0.5,1.0)
    \endpspicture
    + 
   \pspicture[0.4](1.0,1.0)
%  \pspicture[shift=-0.4](1.0,1.0)
    \psframe[linewidth=0.5pt](0,0)(1.0,1.0)
    \psbezier[linewidth=1.0pt](0.25,0)(0.25,0.5)(0.75,0.5)(0.75,1.0)
    \psbezier[linewidth=1.0pt](0.5,0)(0.5,0.25)(0.75,0.25)(0.75,0)
    \psbezier[linewidth=1.0pt](0.5,1.0)(0.5,0.75)(0.25,0.75)(0.25,1.0)
    \endpspicture
    \:\right)
  \\
  -  \frac{d}{d^2-1}
  \left(\: 
 \pspicture[0.4](1.0,1.0)
%  \pspicture[shift=-0.4](1.0,1.0)
    \psframe[linewidth=0.5pt](0,0)(1.0,1.0)
    \psbezier[linewidth=1.0pt](0.25,0)(0.25,0.25)(0.5,0.25)(0.5,0)
    \psbezier[linewidth=1.0pt](0.75,0)(0.75,0.5)(0.75,0.5)(0.75,1.0)
    \psbezier[linewidth=1.0pt](0.5,1.0)(0.5,0.75)(0.25,0.75)(0.25,1.0)
    \endpspicture
    + 
 \pspicture[0.4](1.0,1.0)
%  \pspicture[shift=-0.4](1.0,1.0)
    \psframe[linewidth=0.5pt](0,0)(1.0,1.0)
    \psbezier[linewidth=1.0pt](0.25,0)(0.25,0.5)(0.25,0.5)(0.25,1.0)
    \psbezier[linewidth=1.0pt](0.5,0)(0.5,0.25)(0.75,0.25)(0.75,0)
    \psbezier[linewidth=1.0pt](0.75,1.0)(0.75,0.75)(0.5,0.75)(0.5,1.0)
    \endpspicture
    \:\right)
  = 0
\end{multline}
$k=3$:
\begin{multline}
  \label{eq:JW3}
  \pspicture[0.4](0.9,0.9)
%  \pspicture[shift=-0.4](0.9,0.9)
  \scalebox{0.9}{
    \psframe[linewidth=0.5pt](0,0)(1.0,1.0)
    \psbezier[linewidth=1.0pt](0.2,0)(0.2,0.5)(0.2,0.5)(0.2,1.0)
    \psbezier[linewidth=1.0pt](0.4,0)(0.4,0.5)(0.4,0.5)(0.4,1.0)
    \psbezier[linewidth=1.0pt](0.6,0)(0.6,0.5)(0.6,0.5)(0.6,1.0)
    \psbezier[linewidth=1.0pt](0.8,0)(0.8,0.5)(0.8,0.5)(0.8,1.0)
  }
  \endpspicture
  - \frac{d}{d^2-2}
  \pspicture[0.4](0.9,0.9)
%  \pspicture[shift=-0.4](0.9,0.9)
  \scalebox{0.9}{
    \psframe[linewidth=0.5pt](0,0)(1.0,1.0)
    \psbezier[linewidth=1.0pt](0.2,0)(0.2,0.5)(0.2,0.5)(0.2,1.0)
    \psbezier[linewidth=1.0pt](0.4,0)(0.4,0.2)(0.6,0.2)(0.6,0)
    \psbezier[linewidth=1.0pt](0.8,0)(0.8,0.5)(0.8,0.5)(0.8,1.0)
    \psbezier[linewidth=1.0pt](0.6,1.0)(0.6,0.8)(0.4,0.8)(0.4,1.0)
  }
  \endpspicture
  - \frac{d^2-1}{d^3-2d}
  \left(\: 
  \pspicture[0.4](0.9,0.9)
%  \pspicture[shift=-0.4](0.9,0.9)
    \scalebox{0.9}{
      \psframe[linewidth=0.5pt](0,0)(1.0,1.0)
      \psbezier[linewidth=1.0pt](0.2,0)(0.2,0.2)(0.4,0.2)(0.4,0)
      \psbezier[linewidth=1.0pt](0.6,0)(0.6,0.5)(0.6,0.5)(0.6,1.0)
      \psbezier[linewidth=1.0pt](0.8,0)(0.8,0.5)(0.8,0.5)(0.8,1.0)
      \psbezier[linewidth=1.0pt](0.4,1.0)(0.4,0.8)(0.2,0.8)(0.2,1.0)
    }
    \endpspicture
    + 
   \pspicture[0.4](0.9,0.9)
%  \pspicture[shift=-0.4](0.9,0.9)
    \scalebox{0.9}{
      \psframe[linewidth=0.5pt](0,0)(1.0,1.0)
      \psbezier[linewidth=1.0pt](0.2,0)(0.2,0.5)(0.2,0.5)(0.2,1.0)
      \psbezier[linewidth=1.0pt](0.4,0)(0.4,0.5)(0.4,0.5)(0.4,1.0)
      \psbezier[linewidth=1.0pt](0.6,0)(0.6,0.2)(0.8,0.2)(0.8,0)
      \psbezier[linewidth=1.0pt](0.8,1.0)(0.8,0.8)(0.6,0.8)(0.6,1.0)
    }
    \endpspicture
    \:\right)
  \\
  + \frac{1}{d^2-2}
  \left(\: 
  \pspicture[0.4](0.9,0.9)
%  \pspicture[shift=-0.4](0.9,0.9)
    \scalebox{0.9}{
      \psframe[linewidth=0.5pt](0,0)(1.0,1.0)
      \psbezier[linewidth=1.0pt](0.2,0)(0.2,0.2)(0.4,0.2)(0.4,0)
      \psbezier[linewidth=1.0pt](0.6,0)(0.6,0.5)(0.2,0.5)(0.2,1.0)
      \psbezier[linewidth=1.0pt](0.8,0)(0.8,0.5)(0.8,0.5)(0.8,1.0)
      \psbezier[linewidth=1.0pt](0.6,1.0)(0.6,0.8)(0.4,0.8)(0.4,1.0)
    }
    \endpspicture
    + 
  \pspicture[0.4](0.9,0.9)
%  \pspicture[shift=-0.4](0.9,0.9)
    \scalebox{0.9}{
      \psframe[linewidth=0.5pt](0,0)(1.0,1.0)
      \psbezier[linewidth=1.0pt](0.2,0)(0.2,0.5)(0.6,0.5)(0.6,1.0)
      \psbezier[linewidth=1.0pt](0.4,0)(0.4,0.2)(0.6,0.2)(0.6,0)
      \psbezier[linewidth=1.0pt](0.8,0)(0.8,0.5)(0.8,0.5)(0.8,1.0)
      \psbezier[linewidth=1.0pt](0.4,1.0)(0.4,0.8)(0.2,0.8)(0.2,1.0)
    }
    \endpspicture
    + 
  \pspicture[0.4](0.9,0.9)
%  \pspicture[shift=-0.4](0.9,0.9)
    \scalebox{0.9}{
        \psframe[linewidth=0.5pt](0,0)(1.0,1.0)
        \psbezier[linewidth=1.0pt](0.2,0)(0.2,0.5)(0.2,0.5)(0.2,1.0)
        \psbezier[linewidth=1.0pt](0.4,0)(0.4,0.2)(0.6,0.2)(0.6,0)
        \psbezier[linewidth=1.0pt](0.8,0)(0.8,0.5)(0.4,0.5)(0.4,1.0)
        \psbezier[linewidth=1.0pt](0.8,1.0)(0.8,0.8)(0.6,0.8)(0.6,1.0)
      }
      \endpspicture
      + 
  \pspicture[0.4](0.9,0.9)
%  \pspicture[shift=-0.4](0.9,0.9)
      \scalebox{0.9}{
        \psframe[linewidth=0.5pt](0,0)(1.0,1.0)
        \psbezier[linewidth=1.0pt](0.2,0)(0.2,0.5)(0.2,0.5)(0.2,1.0)
        \psbezier[linewidth=1.0pt](0.4,0)(0.4,0.5)(0.8,0.5)(0.8,1.0)
        \psbezier[linewidth=1.0pt](0.6,0)(0.6,0.2)(0.8,0.2)(0.8,0)
        \psbezier[linewidth=1.0pt](0.6,1.0)(0.6,0.8)(0.4,0.8)(0.4,1.0)
      }
      \endpspicture
      \:\right)
    \\
    - \frac{1}{d^3-2d}
    \left(\: 
  \pspicture[0.4](0.9,0.9)
%  \pspicture[shift=-0.4](0.9,0.9)
      \scalebox{0.9}{
        \psframe[linewidth=0.5pt](0,0)(1.0,1.0)
        \psbezier[linewidth=1.0pt](0.2,0)(0.2,0.2)(0.4,0.2)(0.4,0)
        \psbezier[linewidth=1.0pt](0.6,0)(0.6,0.5)(0.2,0.5)(0.2,1.0)
        \psbezier[linewidth=1.0pt](0.8,0)(0.8,0.5)(0.4,0.5)(0.4,1.0)
        \psbezier[linewidth=1.0pt](0.8,1.0)(0.8,0.8)(0.6,0.8)(0.6,1.0)
      }
      \endpspicture
      + 
  \pspicture[0.4](0.9,0.9)
%  \pspicture[shift=-0.4](0.9,0.9)
      \scalebox{0.9}{
        \psframe[linewidth=0.5pt](0,0)(1.0,1.0)
        \psbezier[linewidth=1.0pt](0.2,0)(0.2,0.5)(0.6,0.5)(0.6,1.0)
        \psbezier[linewidth=1.0pt](0.4,0)(0.4,0.5)(0.8,0.5)(0.8,1.0)
        \psbezier[linewidth=1.0pt](0.6,0)(0.6,0.2)(0.8,0.2)(0.8,0)
        \psbezier[linewidth=1.0pt](0.4,1.0)(0.4,0.8)(0.2,0.8)(0.2,1.0)
      }
      \endpspicture
      \:\right)   + \frac{d^2}{d^4-3d^2+2}
    \pspicture[0.4](0.9,0.9)
%  \pspicture[shift=-0.4](0.9,0.9)
    \scalebox{0.9}{
      \psframe[linewidth=0.5pt](0,0)(1.0,1.0)
      \psbezier[linewidth=1.0pt](0.2,0)(0.2,0.2)(0.4,0.2)(0.4,0)
      \psbezier[linewidth=1.0pt](0.6,0)(0.6,0.2)(0.8,0.2)(0.8,0)
      \psbezier[linewidth=1.0pt](0.8,1.0)(0.8,0.8)(0.6,0.8)(0.6,1.0)
      \psbezier[linewidth=1.0pt](0.4,1.0)(0.4,0.8)(0.2,0.8)(0.2,1.0)
    }
    \endpspicture
    \\
    - \frac{d}{d^4-3d^2+2}
    \left(\: 
    \pspicture[0.4](0.9,0.9)
%  \pspicture[shift=-0.4](0.9,0.9)
      \scalebox{0.9}{
        \psframe[linewidth=0.5pt](0,0)(1.0,1.0)
        \psbezier[linewidth=1.0pt](0.2,0)(0.2,0.2)(0.4,0.2)(0.4,0)
        \psbezier[linewidth=1.0pt](0.6,0)(0.6,0.2)(0.8,0.2)(0.8,0)
        \psbezier[linewidth=1.0pt](0.8,1.0)(0.8,0.4)(0.2,0.4)(0.2,1.0)
        \psbezier[linewidth=1.0pt](0.6,1.0)(0.6,0.8)(0.4,0.8)(0.4,1.0)
      }
      \endpspicture
      + 
  \pspicture[0.4](0.9,0.9)
%  \pspicture[shift=-0.4](0.9,0.9)
      \scalebox{0.9}{
        \psframe[linewidth=0.5pt](0,0)(1.0,1.0)
        \psbezier[linewidth=1.0pt](0.2,0)(0.2,0.6)(0.8,0.6)(0.8,0)
        \psbezier[linewidth=1.0pt](0.4,0)(0.4,0.2)(0.6,0.2)(0.6,0)
        \psbezier[linewidth=1.0pt](0.8,1.0)(0.8,0.8)(0.6,0.8)(0.6,1.0)
        \psbezier[linewidth=1.0pt](0.4,1.0)(0.4,0.8)(0.2,0.8)(0.2,1.0)
}
\endpspicture
\:\right) + \frac{1}{d^4-3d^2+2}
 \pspicture[0.4](0.9,0.9)
%  \pspicture[shift=-0.4](0.9,0.9)
\scalebox{0.9}{
  \psframe[linewidth=0.5pt](0,0)(1.0,1.0)
  \psbezier[linewidth=1.0pt](0.2,0)(0.2,0.6)(0.8,0.6)(0.8,0)
  \psbezier[linewidth=1.0pt](0.4,0)(0.4,0.2)(0.6,0.2)(0.6,0)
  \psbezier[linewidth=1.0pt](0.8,1.0)(0.8,0.4)(0.2,0.4)(0.2,1.0)
  \psbezier[linewidth=1.0pt](0.6,1.0)(0.6,0.8)(0.4,0.8)(0.4,1.0)
}
\endpspicture
\\
=0,
\end{multline}

see Ref.~\onlinecite{Kauffman94} for a recursive formula. These
relations define a finite dimensional Hilbert space $V_d (Y) \subset
\overline{V}_d (Y) \subset \mathbb{C}^{[S]} \subset \mathbb{C}^S$.

In Refs.~\onlinecite{Freedman03,Freedman04a} it is explained that
$V_d(Y)$ is the Hilbert spaces for doubled $SU(2)_k$ Chern-Simons
theory on $Y$.  It has been argued \cite{Freedman05b} that a
Hamiltonian with a ground state manifold corresponding to
$\overline{V}_d$ is perched at a phase transition.  When perturbed,
(infinitesimally for $k=1$ or $2$ and under a larger deformation for
$k\geq 3$) it will go into a topological phase with low-energy Hilbert
space $V_d$, i.e. into a phase described by doubled $SU(2)_k$
Chern-Simons theory.  Very briefly, the Hilbert space of a TQFT such
as doubled $SU(2)_k$ Chern-Simons theory, can always be defined as the
joint null space of commuting local projectors\footnote{See
  Ref.~\onlinecite{Freedman01} where private communication with
  A.~Kitaev and G.~Kuperberg is referenced. The required family of
  commuting projectors is easily derived in the Turaev-Viro approach
  \cite{Turaev} by writing surface$\times$interval, $\Sigma \times I
  =$ handle-body union 2-handles. The disjoint attaching curves of the
  2-handles yield the commuting projectors.}, implying the existence
of a local Hamiltonian with a spectral gap in the thermodynamic limit.
Once a Hamiltonian $H_d$ has imposed $d-$isotopy, i.e.  $GSM(H_d )=
\overline{V}_d$, an extensive degeneracy has been created; the only
\emph{local} way of lifting this extensive degeneracy (to a finite
degeneracy) without creating frustration \footnote{Here we use the
  term ``frustration'' in reference to a Hamiltonian which can be
  written as a sum of projectors yet has no zero-modes. In this sense,
  new terms breaking the extensive degeneracy down to a crystalline
  structure will introduce frustration.} is to add the Jones-Wenzl
projector to such a Hamiltonian.

\section{Microscopic Lattice Models}
\label{sec:micro-models}

We now briefly review some examples of microscopic Hamiltonians whose
ground state(s) are described by $d$-isotopy. These Hamiltonians may
not be particularly realistic (e.g. the two spin Hamiltonians
presented here do not conserve the total spin), but they have the
advantage of being relatively simple local Hamiltonians with the
desired ground states. These ground states have a nice property that their
square-norms are in one-to-one correspondence to the partition functions of
known statistical mechanical models -- a so called \emph{plasma analogy}
\cite{Laughlin83} \footnote{We stress that this is a term-by-term
correspondence in the sense that, given the appropriate basis, the squared
amplitudes of the basis states comprising the ground state are identical to
the Gibbs weights of the corresponding statistical mechanical states (up to
an overall constant).} This approach has proved useful in studying many
frustrated quantum models and, in particular, those with topological and
quasi-topological orders \cite{Rokhsar88,Moessner01a,Misguich02,Henley97,
Henley04,Castelnovo05a, Castelnovo05b}

The first example was presented in Ref.~\onlinecite{Freedman05b} and
was inspired by Kitaev's model \cite{Kitaev97}.  The model is defined
on a honeycomb lattice, with the elementary degrees of freedom being
$s=1/2$ spins situated on its links (alternatively, one can think of
these spins as occupying the sites of a kagom\'{e} lattice, but the former
description lends itself nicer to a loop representation). The
Hamiltonian is given by
\begin{multline}
\label{eq:d-isotopy-Ham}
H_{d}^{(1)} =  {\sum_v} \biggl(1+{\prod_{i\in{\cal N}(v)}}{\sigma^z_i}\biggr)\\
+ {\sum_p} \biggl( \frac{1}{d^2}{\left({F^0_p}\right)^\dagger}{F^0_p} +
{F^0_p}{\left({F^0_p}\right)^\dagger}
- \frac{1}{d}{F^0_p} - \frac{1}{d}{\left({F^0_p}\right)^\dagger}\\
+ {\left({F^1_p}\right)^\dagger}{F^1_p} + {F^1_p}{\left({F^1_p}\right)^\dagger}
- {F^1_p} - {\left({F^1_p}\right)^\dagger}\\
+ {\left({F^2_p}\right)^\dagger}{F^2_p} + {F^2_p}{\left({F^2_p}\right)^\dagger}
-  {F^2_p} - {\left({F^2_p}\right)^\dagger}\\
+{\left({F^3_p}\right)^\dagger}{F^3_p} + {F^3_p}{\left({F^3_p}\right)^\dagger}
- {F^3_p} - {\left({F^3_p}\right)^\dagger}
\biggr)
\end{multline}
where ${\cal N}(v)$ is the set of 3 links neighboring vertex $v$, and
\begin{eqnarray}
{F^0_p} &=& {\sigma^-_1}{\sigma^-_2}{\sigma^-_3}{\sigma^-_4}{\sigma^-_5}{\sigma^-_6}\cr
{F^1_p} &=& {\sigma^+_1}{\sigma^-_2}{\sigma^-_3}{\sigma^-_4}{\sigma^-_5}{\sigma^-_6}
+ \mbox{cyclic perm.}\cr
{F^2_p} &=& {\sigma^+_1}{\sigma^+_2}{\sigma^-_3}{\sigma^-_4}{\sigma^-_5}{\sigma^-_6}
+ \mbox{cyclic perm.}\cr
{F^3_p} &=& {\sigma^+_1}{\sigma^+_2}{\sigma^+_3}{\sigma^-_4}{\sigma^-_5}{\sigma^-_6}
+ \mbox{cyclic perm.}
\end{eqnarray}
Here, $1,2,\ldots,6$ label the six edges of
plaquette $p$.  This Hamiltonian is a sum of projection operators with
positive coefficients and, therefore, is positive-definite.  The
eigenvalues of the first term in (\ref{eq:d-isotopy-Ham}) are $2, 0$,
corresponding, respectively, to whether there is an even or odd number
of ${\sigma^z}=-1$ spins neighboring this vertex. When eigenvalue zero
is obtained at every vertex, the ${\sigma^z}=1$ links form loops.
Hence, the zero-energy subspace of the first term is spanned by all
configurations of multiloops (on the honeycomb lattice, they cannot
cross).

\begin{figure}[hbt]
\includegraphics[width=2.5in]{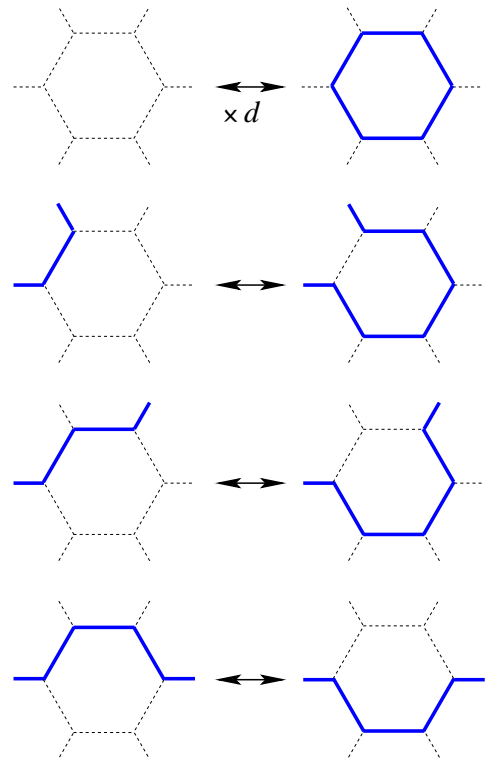}
\caption{The action of the terms in the last four lines of the
  Hamiltonian~(\ref{eq:d-isotopy-Ham}) represented graphically. The
  solid bonds correspond to the up-spins. Notice that the application
  of the Hamiltonian to any of the above plaquette configurations
  results in a superposition of this configuration and its counterpart
  with the appropriate amplitudes. The relative amplitudes of these
  configurations in the ground (0 eigenvalue) state correspond to the
  ``d''-isotopy rules described in the Introduction.}
\label{fig:honeycomb-d-iso}
\end{figure}

The term on the second line of (\ref{eq:d-isotopy-Ham}) is a
projection operator which annihilates a state $|\Psi\rangle$ if the
amplitude for all of the spins on a given plaquette to be up is a
factor of $d$ times the amplitude for them to all be down, i.e. if the
amplitude for a configuration with a small loop encircling a single
plaquette is a factor of $d$ times the amplitude for an otherwise
identical configuration without the small loop.  The other three lines
of the Hamiltonian vanish on a state $|\Psi\rangle$ if it accords the
same value to a configuration if a loop is deformed to enclose an
additional plaquette -- in other words, these terms enforce the usual
isotopy relations. These terms are graphically represented in
Fig.~\ref{fig:honeycomb-d-iso}.

%%%%%%%%%%%%%%%%%%%%%%%%%%%%%%%%%%%%%%%%%%%%%%%%%%%%%%%%%%%
%%%%%%%%%%%%%%%%%%%%%%%%%%%%%%%%%%%%%%%%%%%%%%%%%%%%%%%%%%%%

The second example was presented in Ref.~\onlinecite{Freedman03},
motivated by the connection to a classical statistical mechanical
model, namely the self-dual Potts model (this connection will be
explored in the next section). The model is defined on a square
lattice; the degrees of freedom are once again $s=1/2$ spins situated
on its links. The Hamiltonian is given by
\begin{multline}
  \label{eq:Mike-d-isotopy}
  H_d^{(2)} = \sum_{\pmb{\square}} \left(|3\rangle - \frac{1}{d} |4\rangle
  \right) \left(\langle 3|
    - \frac{1}{d} \langle 4| \right)\\
  + \sum_{\pmb{+}} \left(|\widehat{1}\rangle - \frac{1}{d}
    |\widehat{0}\rangle \right)\left( \langle \widehat{1}| -
    \frac{1}{d} \langle \widehat{0}|\right)
\end{multline}
in the notation of Ref.~\onlinecite{Freedman03}. The sum in
Eq.~(\ref{eq:Mike-d-isotopy}) is taken over all elementary plaquettes
of the lattice, $|3\rangle$ is a state with (any) three up-spins
around a given plaquette, $|4\rangle$ has all four spins up. The
second sum is taken over all vertices; $|\widehat{1}\rangle$ and
$|\widehat{0}\rangle$ correspond to a single spin or no spins up
around a given vertex. Notice that this Hamiltonian is self-dual under
flipping all spins and going to the dual lattice.

\begin{figure}[hbt]
\includegraphics[width=0.5\columnwidth]{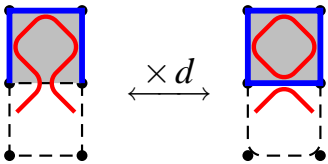}
\caption{The action of the first (plaquette) term of
  Hamiltonian~(\ref{eq:Mike-d-isotopy}) on a shaded plaquette. A small
  loop is created or annihilated (with the appropriate amplitude) by
  being merged with another loop.}
\label{fig:Mikes-d-iso}
\end{figure}

The loops are now defined on the \emph{surrounding} (or midpoint)
lattice, i.e. the lattice obtained by connecting the mid-points of
adjacent edges. One should think of placing a double-sided mirror
along a bond whose spin is up and placing a mirror along a dual bond
if the spin is down. Loops are formed by propagating light in this
labyrinth of mirrors. The action of the first term
Eq.~(\ref{eq:Mike-d-isotopy}) is demonstrated in
Fig.~\ref{fig:Mikes-d-iso}. The action of the second (vertex) term is
completely analogous due to the aforementioned duality.

%%%%%%%%%%%%%%%%%%%%%%%%%%%%%%%%%%%%%%%%%%%%%%%%%%%%%%%%%%%
%%%%%%%%%%%%%%%%%%%%%%%%%%%%%%%%%%%%%%%%%%%%%%%%%%%%%%%%%%%%

Finally, a reader interested in a more realistic Hamiltonian is
referred to [\onlinecite{FNS03b,Freedman05a}] where a possibility of
finding a $d$-isotopy ground state(s) in an extended Hubbard model is
discussed. We shall refrain from reviewing that construction here due
to its complexity, unnecessary for the purpose of this paper.

Let us turn to the common features of the above Hamiltonians. Both the
terms on the final four lines of (\ref{eq:d-isotopy-Ham}) and the
plaquette and vertex terms of (\ref{eq:Mike-d-isotopy}) do not commute
with each other. However, they are compatible in the sense that each
of these terms annihilates the following state\begin{equation}
\label{eq:ground-state}
|{\Psi_0}\rangle = {\sum_{\{\alpha\}}} d^{\ell(\alpha)}\,|\alpha\rangle.
\end{equation}
which is a superposition of all configurations $\alpha$ of multiloops
weighted by a factor of $d$ to the number of loops $\ell(\alpha)$.

Notice one important difference between these examples: by
construction, in the second case the loops are fully packed: every
edge of a surrounding lattice is traversed by a loop. Because of this
constraint, the notion of simple isotopy is meaningless here; the
multiloops, however, can fluctuate by ``ejecting'' and ``absorbing''
small loops.  On an $L\times L$ torus, the ground state degeneracy is
$\sim L^2$ because the Hamiltonian does not mix states
$|\alpha\rangle$ with different winding numbers \footnote{In a
  fully-packed case, there is an additional geometric constraint
  limiting these winding numbers to either only even or only odd,
  depending on the size of the torus}. The different ground states are
given by (\ref{eq:ground-state}) but with the sum over $\alpha$
restricted to a single topological class. Notice that at least in the
first example, the Hamiltonian dynamics is ergodic within a given
topological sector; this is particularly trivial to see for the zero
winding case since every loop configuration can be reduced to a
configuration with no loops by applying the moves depicted in
Fig.~\ref{fig:honeycomb-d-iso} to first shrink the loops to a single
plaquette and then to annihilate them.

In the second model, ergodicity is less obvious since the loop model
is fully-packed and thus reducing any given configuration to a state
with no loops is impossible. Moreover, it has been observed in
Ref.~\onlinecite{Freedman03} that for a non-zero winding number for
certain finite tori there are configurations that are not connected by
the the ``moves'' of the Hamiltonian (\ref{eq:Mike-d-isotopy}). For the
zero winding number, however, it can be shown that all configurations
are connected to the state with all spins down, which in turn
translates into a maximum possible number of loops, one per each
plaquette of a dual lattice. Hence the Hamiltonian is ergodic in the
zero winding sector, and this is the sector we will be concerned with
for the remainder of the paper. (Ergodicity is important because the
twisted states we propose should have nothing to do with degeneracies
associated with possible non-ergodicity.)

More generally, the ground state on any genus $g\geq 1$ surface is
infinitely degenerate in the thermodynamic limit.  As we saw in the
previous section if $d=\pm1$ or $d=\pm\sqrt{2}$, there is a
Jones-Wenzl projector which also annihilates the ground state
(\ref{eq:ground-state}) on a topologically-trivial manifold but mixes
different winding number sectors on higher-genus surfaces.  Hence, in
either model, at $d=\pm 1, \pm \sqrt{2}$, there are two Hamiltonians
which have the same ground state on the sphere (or, equivalently, in
the zero winding sector). The first ones, given by either
Eq.~(\ref{eq:d-isotopy-Ham}) or Eq.~(\ref{eq:Mike-d-isotopy}), have
extensively degenerate ground states on the torus while the other kind
(with the Jones-Wenzl projector added) have finite degeneracy
\cite{Kitaev97}.  The second type leads to a topological phase with an
energy gap for $d=1$ \cite{Kitaev97}, where the resulting model is
exactly soluble, and, we believe, for $d=-1$ (the situation at
$d=\pm\sqrt{2}$ is more subtle and will be addressed later). The
spectrum of the first kind is the main subject of this paper.

While the ground state can be obtained exactly, excited states cannot
because the different operators in both (\ref{eq:d-isotopy-Ham}) and
(\ref{eq:Mike-d-isotopy}) do not commute with each other.  We will use
a variational \emph{ansatz} to show that in the absence of other terms,
such as Jones-Wenzl projectors, the spectra of these Hamiltonians are
gapless in the zero winding sector. But before we can proceed, we need
to address the inherent structure of these ground states in more
details.

\section{Mapping of the Ground-State to a Statistical Mechanics
Problem}

Many properties of the ground-state wavefunction can be obtained by
observing that the norm of the ground state is equal to the partition
function of a classical loop model:
\begin{eqnarray}
\label{eq:sum-over-configs}
\langle {\Psi_0}|{\Psi_0}\rangle = \sum_{\{\alpha\}} {d^{2{\ell(\alpha)}}}
\end{eqnarray}
where $\alpha$ denotes a particular configuration of loops on a
lattice (a ``snapshot'' of multiloops). The specific details of
possible configurations depend on a particular choice of a
Hamiltonian; in what follows we shall consider the cases relevant to
each of the proposed Hamiltonians.

\subsection{Potts model: random clusters and loops}
\label{sec:Potts}

The $q$-state Potts model, originally introduced as a generalization
of the Ising model, is defined by the following classical Hamiltonian:

\begin{equation}
 - \beta \mathcal{H} = J\sum_{<i,j>}{\delta_{\sigma_i,
      \sigma_j}} \,,
  \label{eq:Potts_Hamiltonian}
\end{equation}
where the sum is taken over all pairs of nearest neighbors and
$\sigma_i$ is a discrete ``spin'' variable that can take on $q$
different values, e.g.\ $\sigma_i = 1, \ldots, q$.

We now review the the Fortuin--Kasteleyn (\emph{AKA} random cluster)
representation for the $q$-state Potts model and underline its basic
properties.

Given the Hamiltonian \ref{eq:Potts_Hamiltonian}, the partition
function can then be written as
\begin{multline}
    \label{eq:Potts_partition}
      Z_{\text{Potts}}
  = \sum_{\{ \sigma\}}\mathrm{e}^{ - \beta \mathcal{H}} =\sum_{\{
          \sigma\}} \prod_{<i,j>}\mathrm{e}^{J \delta_{\sigma_i,
              \sigma_j}} \\
       = \sum_{\{
          \sigma\}}\prod_{<i,j>}\left[1+\left(\mathrm{e}^{J}-1\right)
          \delta_{\sigma_i, \sigma_j}\right]  \\
       = \sum_{\{ \sigma\}}\prod_{<i,j>}\left[1+ v\,
          \delta_{\sigma_i, \sigma_j}\right]\,,
\end{multline}
with $v\equiv \mathrm{e}^{J}-1$.

The next step is to expand the product in
Eq.~(\ref{eq:Potts_partition}).  Every pair of nearest neighbors
contributes a factor of either 1 or $v$ to each of the resulting
terms, with the latter possibility available \emph{only} if the
neighboring spins agree ($\sigma_i = \sigma_j$). Therefore every such
term has a simple graphical representation: a bond is occupied if the
pair of spins it connects contributes a factor of $v$ into a given
term, and it is left vacant otherwise. Since a bond can be placed
between the sites \emph{only} if their spins ``agree'', all spins
belonging to the same bond cluster must have the same value.

The partition function thus becomes
\begin{equation}
    \label{eq:Potts_partition_expanded}
    Z_{\text{Potts}}  
  = \sum_{\{ \sigma\}}\prod_{<i,j>}\left[1+ v\,
    \delta_{\sigma_i, \sigma_j}\right]
  = \sum_{\{ \sigma\}}\sum_{\{ \omega\}} v^{b(\omega)}
    \Delta(\sigma,\omega)\,,
\end{equation}
where $b(\omega)$ is the total number of occupied bonds in a bond
configuration $\omega$, and $\Delta(\sigma,\omega)$ is the appropriate
collection of Kronecker delta-symbols that enforces the ``agreement''
between the spin and the bond configurations.

The next step is to change the order of summation in
Eq.~(\ref{eq:Potts_partition_expanded}) (which is fine as long as the
lattice is finite) and then sum over all spin configurations $\sigma$.
The only constraint on the spin variables (for a fixed bond
configuration $\omega$) is the one that has been mentioned earlier:
all connected spins must take on the same value (one of the $q$
possible). Therefore every connected cluster, as well as each isolated
site, contributes a factor of $q$ to the resulting bond weight, and
the partition function becomes
\begin{equation}
  \label{eq:FK_rep1}
  Z_{\text{Potts}}  =\sum_{\{ \omega\}}v^{b(\omega)}q^{c(\omega)} ,
\end{equation}
where $c(\omega)$ is the total number of connected components
(including isolated sites).  The partition function (\ref{eq:FK_rep1})
defines FK (random cluster) representation \cite{FK-72}\footnote{This
  derivation can be made slightly more formal if one defines a joint
  measure on both spin and bond configurations: $\mu(\sigma, \omega) =
  v^{b(\omega)} \Delta(\sigma,\omega)$. This is known as
  Edwards--Sokal measure \cite{ES-88}.  As follows from
  Eq.~(\ref{eq:Potts_partition_expanded}), when traced over both spins
  and bond occupations, it gives the desired partition function. Its
  marginal with respect to bond configurations is the Gibbs weight for
  spins, while its marginal with respect to spin configurations
  defines the weight for random clusters.}

While all previous considerations applied to any number of spatial
dimensions, in 2D the partition function (\ref{eq:FK_rep1}) can also
be written in a way which makes its self-dual property apparent.  In
order to do this, notice that if we think of an unoccupied bond of the
actual lattice as an occupied bond of the dual lattice (therefore
$b^*=B-b$), then every circuit (face) of the actual lattice contains a
dual connected component: $f=c^*$ (here we define $b^*\equiv
b(\omega^*)$, $c^*\equiv c(\omega^*)$ with $\omega^*$ denoting the
configuration of dual bonds).  Therefore, using the Euler relation
(for a planar graph)
\begin{equation}
  \label{eq:Euler}
  f(\omega) = b(\omega) + c(\omega) - N
\end{equation}
with $f(\omega)$ being the number of \emph{circuits} (defined as a
minimum number of bonds that one has to cut in order to make a graph
consist only of \emph{trees} \footnote{It is also known as a
\emph{cyclomatic number} and for the case of a planar graph is equal
 to the number of \emph{finite faces}
\cite{Essam-Fisher-70:graphs}.}) and $N$ being the total number of
sites,  we can rewrite
Eq.~(\ref{eq:FK_rep1}) as
\begin{equation}
    \label{eq:FK_rep3}
    Z_{\text{Potts}}
  =\sum_{\{ \omega\}}v^{f-c+N}q^{c} = v^N \sum_{\{ \omega\}} v^f
      \left(\frac{q}{v}\right)^c = v^N \sum_{\{ \omega\}} v^{c^*}
      \left(\frac{q}{v}\right)^c
\end{equation}
If we ignore the uninteresting analytic prefactor in
Eq.~(\ref{eq:FK_rep3}), we immediately see that the system is
self-dual when $v=q/v$, i.e. $v=\sqrt{q}$.

%%%%%%%%%%%%%%%%%%%%%%%%%%%%%%%%%%%%%%%%%%%%%%%%%%%%%%%%%%%
\begin{figure}[hbt!]
\includegraphics[width=6.0cm]{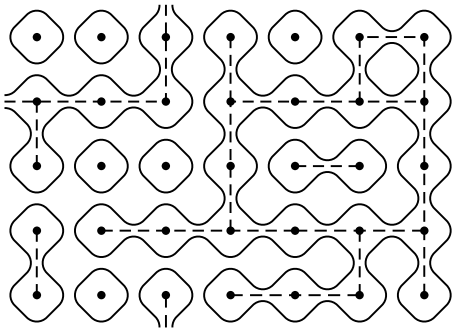}
\caption{A typical cluster configuration for the Potts model
  is shown by dashed lines. Spins belonging to the same cluster take
  the same value, which must be summed over the $q$ possible values,
  as described in the text. Clusters can be represented by loops on
  the surrounding lattice, shown by solid lines. }
\label{fig:Potts-loops}
\end{figure}
%%%%%%%%%%%%%%%%%%%%%%%%%%%%%%%%%%%%%%%%%%%%%%%%%%%%%%%%%%%%

A so-called polygon decomposition \cite{BKW-76} lets us relate random
clusters to a loop gas on the surrounding lattice.
%(vertices of the surrounding lattice are
%the midpoints of the original bonds).
We think of an occupied bond as a double-sided mirror placed at the
site of the surrounding lattice. If a bond is not occupied, then its
dual bond is considered a mirror. Thus every site of the surrounding
lattice gets one of the two possible mirrors. We then use these
mirrors to construct paths as shown in figure~\ref{fig:Potts-loops}.
Since these paths have no sources or sinks, they always form loops
that either surround the clusters or are contained inside clusters (in
the latter case, the loops surround \emph{dual} clusters). The number
of loops $\ell$ is then given by $\ell=c+c^*$. If $v=\sqrt{q}$ -- i.e.
if the Potts model is at its self-dual point, then
\begin{eqnarray}
\label{eq:Potts-loops}
Z_{\text{Self-Dual}} = \sum_{\{ \alpha \}}
{\left(\sqrt{q}\right)^{\ell(\alpha)}}.
\end{eqnarray}
where the sum is taken over all fully-packed loop configurations on
the surrounding lattice. Clearly, with the choice of $q=d^4$, the
partition function (\ref{eq:Potts-loops}) becomes the norm of the
ground state $\langle {\Psi_0}|{\Psi_0}\rangle$ of the quantum
Hamiltonian $H_{d}^{(2)}$.

\subsection{Random clusters on a torus}
\label{sec:mikes_thm}

In this section we will take the reader through some mathematical
details in order establish certain properties of  
%For the sake of concreteness, let us focus on the $d$-isotopy
%Hamiltonian corresponding to 
the critical FK model. These properties will be relied upon later, in the
course of demonstrating the central result of this paper. 

For the sake of concreteness, let us focus
the FK representation of the critical $q-$state Potts model ($1 \leq q \leq
4$)
%, $1 \leq d \leq \sqrt{2}$,
on a square lattice with
periodic boundary conditions in both directions, i.e. the torus.
In the case of $q=1$, the statistical mechanical model reduces to
critical bond percolation. In this section we are focusing on the FK clusters
% rather then the loops that surround them; their relation has been
established in Sec.~\ref{sec:Potts}.

Let us begin by proving:

\begin{proposition}
\label{prop_fkcluster}
Fixing $1 \leq q \leq 4$ and $\lambda > 0$ there is an $\epsilon > 0$
so that for all $L$ sufficiently large there is a probability greater than
$\epsilon$ that the largest $\text{FK}_q$ cluster in the $L \times
L$ torus has Euclidian diameter $< \lambda L$.
\end{proposition}

\begin{proof}
The technology for this type of result was discovered and developed
by the 1990's in the context of critical percolation
\cite{FKG-71,ACCN-88,Grimmett-book,CPS-99} but it can be extended to other
critical systems provided
analogs of the Russo--Seymour--Welsh (RSW) inequality on crossing
probabilities for rectangles and of the Fortuin--Kasteleyn--Ginibre (FKG)
inequality (``monotone
events are positively correlated'') hold.  The proof strategy is to
build some {\em wiggly} approximation $G_\text{w}$, lying entirely in the
complement of FK clusters, to a rectilinear grid of scale
(roughly) $\frac{L}{2\lambda}$.  $G_\text{w}$ must not be {\em too} wiggly
as we need the boxes in its complement to have diameter~$< \lambda$ (see 
Fig.~\ref{fig:wiggly-approx}).
Since each FK cluster lies in some such box, the clusters also have
diameter~$< \lambda$.

%%%%%%%%%%%%%%%%%%%%%%%%%%%%%%%%%%%%%%%%%%%%%%%%%%%%%%%%%%%
\begin{figure}[hbt]
\includegraphics[width=0.8\columnwidth]{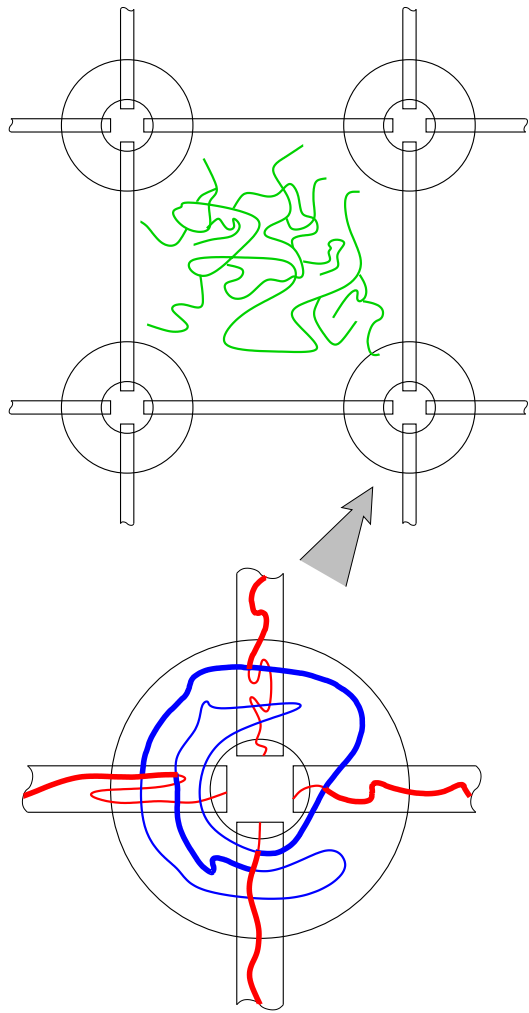}
\caption{Construction of a wiggly approximation $G_\text{w}$ to a rectangular
grid as described in the text. $G_\text{w}$ is built as a subset of the
\emph{dual} clusters; the goal of the construction is to ``trap'' the actual
FK clusters inside $G_\text{w}$, as shown here.}
\label{fig:wiggly-approx}
\end{figure}
%%%%%%%%%%%%%%%%%%%%%%%%%%%%%%%%%%%%%%%%%%%%%%%%%%%%%%%%%%%%

Building the grid $G_\text{w}$ is a random construction and must succeed
with probability $> \epsilon$.  The naive idea for building the grid
is to first condition on no clusters meeting the (roughly)
$\left(\frac{L}{2\lambda}\right)^2$ sites (i.e. $4$-coordinated
vertices) of the desired grid $G_S$.  Then, use RSW to further
condition on cluster-disjoint arcs lying in small rectangles near
the (roughly) $2 \left(\frac{L}{2\lambda}\right)^2$ bonds of $G_S$.
What goes wrong is that the individual sites are simply too small;
in the analogous electromagnetic problem the
capacitance vanishes in the thermodynamic limit.  The resolution is to
consider small annuli enclosing each site and substitute for the
site condition the event that an essential circle percolates around
the annulus (also an RSW result).

Now, elementary planar topology allows bits and pieces of the
annular and rectangular percolation paths to be hooked together to
build the desired $G_\text{w}$ as shown in Fig.~\ref{fig:wiggly-approx}. 
Throughout this construction, it is
essential that these percolation events are at least non-negatively
correlated (we remind the reader that the properties of dual clusters are
identical to those of the actual FK clusters at criticality). This is
precisely what the FKG inequality provides
\cite{FKG-71,ACCN-88}.
\end{proof}

An immediate consequence of proposition \ref{prop_fkcluster} is that
there is a non-zero probability that an arbitrary multi-loop
is in the zero-winding-number sector.  Thus, we may
confine our discussion to that sector and use the measure
on that sector induced by the measure $d^{2\ell(\alpha)}$ on all multi-loops.

Let us pause for some definitions.  Given a loop $a$ in the torus
$T$, let $||a||_x$, or simply $||a||$ for short, be the ``breadth in
the $x$-direction of $\widetilde{a}$'' where $\widetilde{a}$
represents a (any) lift of $a$ up to the universal cover
$\mathbb{R}^2 \rightarrow T^2$.  Recall that the universal cover of
the torus $T^2$ is constructed by unwrapping completely in both the
$x$ and $y$ directions.  Very concretely, we may build the cover
$\mathbb{R}^2$ from our original $L \times L$ square by taking a
copy for each Gaussian integer and gluing these together
(corresponding to Gaussian integers differing by $1$ or $i$) to form
a tiling of the plane.  (Technically, it is the discrete plane
$\mathbb{Z}^2$ that we produce.)  A loop $a$ in $T^2$ may be
regarded as a map from the circle $S^1$ into $T^2$ and a lift
$\widetilde{a}$ is any map to $\mathbb{R}^2$ making the following
diagram commute:

\[\begindc{0}[30]
    \obj(1,1)[A]{$S^1$}
    \obj(3,1)[B]{$T^2$}
    \obj(3,3)[C]{$\mathbb{R}^2$}
    \mor{A}{B}{$a$}
    \mor{C}{B}{}
    \mor{A}{C}{$\widetilde{a}$}[\atleft, \dasharrow]
\enddc\]

Concretely, the image of $\widetilde{a}$ is obtained by taking an
arc in the $L \times L$ square and following its continuation in the
tiling of $\mathbb{R}^2$ until it closes back on itself.  The
assumption that the winding sector is trivial means any such
$\widetilde{a}$ is finite and in fact contains exactly as many bonds
as $a$ contained in $T^2$, i.e. $< 2L^2$.

To return to the definition of $||a||$, we say $||a|| = \max|x_i -
x_j|$, where the maximum is taken over all sites $(x_i, y_i)$ on a
lift $\widetilde{a}$.  Note that any two lifts $\widetilde{a}$ and
$\widetilde{\widetilde{a}}$ are congruent by a translation of
$\mathbb{R}^2$, so $||a||$ is well defined.

Finally, for a multi-loop $\alpha \subset T^2$ we define $||\alpha||
= \max_{a \subset \alpha} ||a||$.  We are now ready to state a
proposition about the diameter of these unwrapped components of
random multi-loops from the trivial sector.

\begin{proposition}
\label{prop_normalpha} In the scaling limit, $||\alpha||$ is almost surely a
bounded
function on the trivial winding sector. More precisely, the probability that
$||\alpha||>rL$
has an upper bound which is independent of $L$.
\end{proposition}

\begin{proof}
We need a geometric lemma.

\begin{lemma}
\label{lemma_torus} Let $T$ be a Euclidian torus of area $1$.  Let
$\gamma : [0,1] \rightarrow T$ be an arc so that the (Euclidian)
distance between lifted endpoints in the universal cover,
$d(\widetilde{\gamma}(0),
\widetilde{\gamma}(1)) = \delta$.  Then there is a nontrivial deck
translation (additive action of a Gaussian integer)
$\widetilde{\gamma}'$ of $\widetilde{\gamma}$ so that
$dist(\widetilde{\gamma}, \widetilde{\gamma}') < \frac{1}{\delta}$.
\end{lemma}

\begin{proof}
Let $X$ be the radius $r$ neighborhood of $\gamma$ in the in the
cover $\mathbb{R}^2$.  Integrating the lengths of slices of $X$
perpendicular to the straight line segment $[\widetilde{\gamma}(0),
\widetilde{\gamma}(1)]$ we see (by Fubini's theorem) that $\text{area}(X) >
2r\delta$ (where area is counted {\em without} multiplicity).  If
$X$ is disjoint from its deck transformations then it descends
one-to-one into the torus, implying $\text{area}(X) < 1$.  Thus, $2r\delta
< 1$ and so $r < \frac{1}{2\delta}$.
\end{proof}

Now apply the lemma to arcs $\gamma$ within an FK cluster $K$ (in
the trivial sector) on $T$.  Since we are in the trivial sector, we
may define $||K||$ exactly as we defined the norm on loops.  By the
lemma, if $||K|| = \delta$, $K$ will come within distance
${1}/{\delta}$ of completing a nontrivial wrapping of the
torus.  An application of the Russo-Seymour-Welsh (RSW) inequality
shows that it is unlikely (vanishing algebraically in ${1}/{\delta}$) to
come close to wrapping $T^2$ but fail to wrap completely.  RSW
amounts to integrating the effect of bringing bonds, which may join large
clusters,
in and out of our FK snapshot.  Such fluctuations
{\em cannot} be created by any local $H_d$ for $q \neq 1$ since the
weight of the update is nonlocal.  Nevertheless, these fluctuations do
preserve the measure $d^{2\ell(\alpha)}$.  So, the fraction of critical,
topologically trivial, $\text{FK}_q$ snapshots ($1 \leq q \leq 4$)
which contain a cluster $K$ with $||K|| > \delta$ is
$\frac{1}{\text{poly}(\delta)}$.
\end{proof}

\subsection{$\text{O}(n)$ loop model}
\label{sec:On}

Another model describing the statistics of loops on a lattice
is the so-called $\text{O}(n)$ loop model \cite{Domany-81}. Let us begin by
defining the following $\text{O}(n)$ spin model on some (finite)
lattice by the following \emph{partition function}:
\begin{equation}
  \label{eq:O(n)_part}
  Z_{\text{O}(n)}(x) = \int {\prod_i}
  {\frac{d \hat{S}_i}{\Omega_n}}\,\prod_{\langle i,j\rangle}
  (1+x\,{\mathbf{S}_i}\cdot {\mathbf{S}_j})
\end{equation}
with $\mathbf{S}_i \in {\mathbb R}^n$, $\left|{\mathbf{S}_i}\right| =
1$ and $\Omega_n$ is the $n$-dimensional solid angle.  In order to
obtain the loop model, we write $\mathbf{S}_i \cdot \mathbf{S}_j =
S_i^{(1)} S_j^{(1)} + \ldots + S_i^{(n)} S_j^{(n)}$ and define $n$
different colors (each of which will be associated with a specific
component of the $O(n)$-spins).  Multiplying out all terms in
Eq.~(\ref{eq:O(n)_part}), we have $n$ choices for each bond plus a
possibility of a vacant bond.  Thus, the various terms are represented
by an $n$-colored bond configuration: $\mathcal{G} = \mathcal{G}_1,
\ldots \mathcal{G}_n$ with $\mathcal{G}_\ell$ denoting those bonds
where the term $S_i^{(\ell)} S_j^{(\ell)}$ has been selected. Clearly,
the various $\mathcal{G}_\ell$'s are pairwise (bond) disjoint. Thus,
for each $\mathcal{G}$ we obtain the weight
\begin{equation}
  W_\mathcal{G} = \mathrm{Tr} \prod_{\langle
    i, j \rangle \in \mathcal{G}_1} x \, S_i^{(1)} S_j^{(1)} \ldots
  \prod_{\langle i, j \rangle \in \mathcal{G}_n} x \, S_i^{(n)}
  S_j^{(n)}\,.
  \label{bond_weight}
\end{equation}
On the basis of elementary symmetry considerations it is clear that
$W_\mathcal{G} \neq 0$ if and only if each vertex houses an even
number (which could be 0) of bonds of each color. Once this
constraint is satisfied, we get an overall factor of
$x^{b(\mathcal{G})}$ -- with $b(\mathcal{G})$ being the total
number of bonds -- times the product of the \emph{vertex factors}
obtained by performing the appropriate $O(n)$ integrals.
% This procedure generates loops of $n$ different colors.
Obviously, these vertex factors depend only on how many different
colors and how many of each of these colors enter each vertex
(i.e.\ not on the particular colors involved nor on the directions
of approach to the vertex). A particularly easy case is that of the
honeycomb lattice where, due to a low coordination number, a maximum of
two bonds of a single color can visit a vertex. The corresponding
vertex factor is then given by
\begin{equation}
\label{eq:O(n)_vert}
\int
{\frac{d \hat{S}_i}{\Omega_n}}\, \left(S_i^{(j)}\right)^2 = \frac{1}{n}
\end{equation}
leading (after summing over all $n$ colors) to the following
expression for the partition function:
\begin{equation}
  \label{eq:O(n)-def}
  Z_{\text{O}(n)}(x) ={\sum_{\{\alpha\}}}
  \left(\frac{x}{n}\right)^{b(\alpha)}\,n^{\ell(\alpha)}
\end{equation}
where $b(\alpha)$ is the total number of occupied bonds (the total
perimeter of all loops) while ${\ell(\alpha)}$ is the total number of
loops. The last factor appears because each loop could be of one of
$n$ colors.  The expression in (\ref{eq:O(n)-def}) is well-defined for
arbitrary $n$ and $x$, so it can be taken as the \emph{definition} of
the $O(n)$ loop model \cite{Nienhuis87}.

We note that the partition function (\ref{eq:O(n)-def}) is
\emph{exactly} the norm of the ground state $\langle
{\Psi_0}|{\Psi_0}\rangle$ of the quantum Hamiltonian $H_{d}^{(1)}$
given by Eq.~(\ref{eq:d-isotopy-Ham}) provided that $x=n$ and $d^2 =
n$.

\subsection{Correlations}
\label{sec:correlations}

Both of these models have a Coulomb gas representation
\cite{Nienhuis87} so that their correlation functions can be obtained
from exponential operators in a Gaussian field theory with a
background charge.  Consider, for instance, the O$(n)$ model.
Precisely the same loop expansion derived in section \ref{sec:On} can
be obtained from an SOS model on the kagom\'{e} lattice. After integrating
out the triangular faces, the resulting SOS model for the heights on
the hexagonal faces is has a low-temperature expansion which is a sum
over domain wall configurations.  The weights of the SOS model (or,
equivalently, 6-vertex model) are such that when a domain wall turns
left, it acquires a factor $({x}/{n})e^{i\chi}$; when it turns
right, a factor $({x}/{n})e^{-i\chi}$.  Since a loop will only close
if the difference between the number of right terms and the number of
left turns is $\pm 6$, every closed loop receives a factor
${\left({x}/{n}\right)^b} e^{\pm 6 i\chi}$ where $b$ is the length
of the loop. Summing over both orientations of the loop, we obtain
$2{\left({x}/{n}\right)^b} \cos{ 6\chi}$. Hence, this is
equivalent to the O$(n)$ loop model for $n<2$ if we take $n=2\cos{
  6\chi}$.  When $x>{x_c}=n/\sqrt{2+\sqrt{2-n}}$, this model is in its
low-temperature phase, which is critical.

The SOS model is a Coulomb gas with coupling $g=1-{6\chi}/{\pi}$
and background charge $-2(1-g)$ (which ensures the correct phase
factor for each turn of a loop). Consider the $\langle {\bf S_i}\cdot
{\bf S_j}\rangle$ correlation function in the O$(n)$ model. Every
configuration which gives a non-zero contribution must have one curve
which does not close into a loop but has endpoints at spins at $i$ and
$j$. In Coulomb gas language, they
have magnetic charges $\pm 1/2$.  In addition, they must each also
have electric charge $1-g$. Together with their magnetic charge, this
will ensure that a factor $e^{\pm 6i\chi}$ arises whenever the curve
winds around either point (as required by the SOS vertex rules).  This
electric charge also cancels the background charge.  Thus, the Coulomb
gas operators corresponding to O$(n)$ spin operators have electric and
magnetic charges $(1-g,\pm 1/2)$.  The exponent associated with a
correlation function between field with electric and magnetic charges
$({e_1},{m_1})$ and $({e_2},{m_2})$ is
$x_{{e_1},{m_1};{e_2},{m_2}}=-{{e_1}{e_2}}/{2g}
-{g{m_1}{m_2}}/{2}$.  Hence, the O$(n)$ spin-spin correlation
function has the power-law decay \cite{Nienhuis87},
\begin{equation}
\label{O(n)-spin-spin}
\left\langle \mathbf{S}(r) \cdot \mathbf{S}(0)\right\rangle \sim
\frac{1}{r^{x_M}},
\end{equation}
where ${x_M}=\frac{1}{4}\,g - \frac{1}{g}(1-g)^2$.

The loops which surround random clusters as described in
Sec.~\ref{sec:Potts} in the self-dual critical $q$-state Potts model
can also be mapped onto a 6-vertex model (and, therefore, a Coulomb
gas) in a similar fashion.  Precisely the same exponents are obtained.

Although, they are crucial to our proof of gaplessness, these
correlation functions are not the ones of direct physical interest.
Physical correlation functions have a rather different behavior.
Consider equal-time correlations between \emph{quantum} spins such as
$\langle{\sigma^{z}_i}{\sigma^{z}_j}\rangle$ in the original quantum
models (\ref{eq:d-isotopy-Ham},\ref{eq:Mike-d-isotopy}). As we now
argue, they are \emph{short-ranged in space}.  In the first model, the
equal-time $\langle{\sigma^{z}_i}{\sigma^{z}_j}\rangle$ correlation
function is related to the probability that a loop passes through $i$
and a loop which may or may not be distinct passes through $j$. Such
correlation functions vanish in the $O(n)$ loop models. Analogously,
in the second case, this correlation function is related to the
probability that the two corresponding bonds are parts of clusters,
but not necessarily in the same cluster in the related $q$-state Potts
model. Such a correlation function once again vanishes.

In Coulomb gas language for the associated statistical mechanical
models, the reason that such correlation functions vanish is that they
are correlation functions of electrically neutral operators, such as
gradients of the height (to which the local loop density corresponds).
Such correlation functions vanish since they do not cancel the
background charge.  (The only exception is a height model, with
central charge $c=1$, for which there is no background charge.
Gradients of the height have power-law correlations.) As we have seen,
algebraic decay is possible for correlation functions of operators
which are charged in the Coulomb gas picture, but these are non-local
in terms of the spins ${\sigma^{z}_i}$ since they measure, for
instance, the probability that two spins ${\sigma^{z}_i}$ and
${\sigma^{z}_j}$ lie on the \emph{same loop}. (At $d=1,\sqrt{2}$, this
can also be seen from the fact that the ground state on the sphere is
the same -- and, therefore, has the same equal-time correlation
functions -- as that of a gapped Hamiltonian \cite{Kitaev97,Turaev92}
which is a sum of local commuting operators and, therefore, has
correlation length zero.)

Thus, the ground state wavefunction of (\ref{eq:d-isotopy-Ham}) has an
underlying power-law long-ranged structure which is apparent in its
loop representation, but it is not manifested in the correlation
functions of local operators ${\sigma^{z}_i}$.  As we will see
momentarily, this long-range structure leads to gapless excitations
for the Hamiltonian (\ref{eq:d-isotopy-Ham}) and, therefore,
long-ranged correlations in time in spite of the lack of long-ranged
correlations in space. We call such a state of matter a
\emph{quasi-topological} critical point.

\section{Low-Energy Excitations}

In spite of the short-ranged nature of equal-time spin-spin
correlation functions and the absence of any conservation laws for
either of the Hamiltonians
(\ref{eq:d-isotopy-Ham},\ref{eq:Mike-d-isotopy}), we can construct a
variational argument that this general type of Hamiltonians is gapless
using the criticality of non-local correlation functions, specifically
the scale-invariant nature of loops.

The general idea of our proof is to produce a ``twisted'' state which is
both orthogonal to the ground state and has a vanishingly small
expectation value of the Hamiltonian, in the general spirit of the
Lieb-Schultz-Mattis theorem for quantum antiferromagnets
\cite{Lieb61,Hastings04a,Hastings05a}.

For the purposes of this theorem we do not need to choose a particular
Hamiltonian, but only require that it satisfy the following necessary
conditions:
\begin{itemize}
\item This is a $d$-isotopy Hamiltonian whose ground state is described by
Eq.~(\ref{eq:ground-state}) with a scale-invariant distribution of
loops in the thermodynamic limit. We make the mathematically
nontrivial assumption, widely accepted in physics, that the
scaling limit exists.  Furthermore, certain probability
functions defined from the limit will be formally differentiated.
These derivatives can easily be replaced by finite difference
quotients, so continuity is, in fact, an adequate assumption. (We
shall make this condition more precise later.)

\item A second important feature of the ground state which we use is
that when viewed as a statistical mechanical ensemble by $\text{Prob}(\psi_i)
=
|\psi_i|^2$, it is, in the scaling limit, a critical system with the
``no large cluster property,'' i.e. on an $L \times L$ square with
periodic boundary conditions (``torus'') $\forall \lambda > 0,
\exists \epsilon > 0$ so that the event ``all loops within the
random multi-loop $\psi_i$ have breadth (as defined is
Section~\ref{sec:mikes_thm}) $< \lambda L$'' occurs with
probability greater than $\epsilon$.  (We have seen that the FK
models for $q \geq 1$ have this property. We also expect this property to hold
for the O($n$) loop models, $1\leq n \leq 2$; proving this would be quite
interesting.)

\item The Hamiltonian is local: all allowed $d$-isotopy moves have finite
range (generalizing our theorem to the case of a quasi-local
Hamiltonian whose terms decay exponentially with increasing range is
straightforward). Without loss of generality, we might as well assume that
the range of all terms is limited to a single lattice plaquette.

\item The terms responsible for the loop dynamics are bounded uniformly in
the size of the system, i.e. for any two multiloops $\alpha$ and
$\beta$, $\left|\langle \alpha| H_d | \beta \rangle \right| < V$. (We
assume that the basis vectors $|\alpha\rangle$ of the Hilbert space of
$H_d$ are orthonormal, i.e.  $\langle \alpha| \beta \rangle = 0$
unless $\alpha$ and $\beta$ are identical multi-loops and $\langle \alpha|
\alpha \rangle = 1$). Notice that we make no assumptions about the
other terms that might be there to enforce the multiloop constraint
-- these can potentially include hard-core interactions.
\end{itemize}
There is little doubt that both of the presented Hamiltonians
satisfy the above conditions.  However, our argument is more
complete in the second case (FK$_q$, $1 \leq q \leq 4$).

%%%%%%%%%%%%%%%%%%%%%%%%%%%%%%%%
%% Mike's stuff extracted %%%%%%
%%%%%%%%%%%%%%%%%%%%%%%%%%%%%%%%

We will now construct our low energy excitation $|{\psi_1}\rangle$ above the
ground state $|{\psi_0}\rangle$.  We normalize so that $\la \psi_0 | \psi_0
\ra = 1$, $\la \psi_0 | H | \psi_0 \ra = 0$, and $\la \psi_1 |
\psi_1 \ra = 1$.  We will construct $|{\psi_1}\rangle$ and then a family of
``harmonics'' $|{\psi_k}\rangle$, all of norm one, and estimate the energy
expectation values $\la \psi_k | H | \psi_k \ra$, $k \geq 1$.  In
constructing $|{\psi_k}\rangle$, we will use the language of the scaling limit
for conceptual simplicity.  It is quite routine, and we leave this
to the reader, to back away from the scaling limit and write
discrete formulae, replacing derivatives
with difference quotients.

We use ${\cal C}$ to denote a configuration (i.e. a multi-loop)
near the scaling limit, i.e. $L \to \infty$, on $T$, the $L \times L$ torus. 
By proposition
\ref{prop_normalpha} we know there is a function $r_{\cal C}$ on
configurations which remains almost surely defined in the scaling
limit: $r_{\cal C} \equiv {||{\cal C}||}/{L}$.  Our assumption is that, in the scaling limit,
the probability $p$ that a configuration
$\cal C$ satisfied $r_{\cal C} \leq r$ is a continuous function $p(r)$.
The probability is computed with respect to the $FK_q$ measure, i.e. with
respect to $|{\psi_0}\rangle$.
As explained earlier, we will formally treat $p$ as differentiable,
writing ${dp}/{dr}$, but this may be treated as a difference
quotient.

The ground state wave function (on the trivial sector) $|{\psi_0}\rangle$
is simply:
\begin{equation}
|{\psi_0}\rangle = Z^{-1/2}{\sum_{\alpha}}d^{\ell(\alpha)}\,|\alpha\rangle
\end{equation}
where the sum is over all multi-loops $\alpha$ in the zero-winding-number
sector and the normalization
$Z={\sum_{\{\alpha\}}} d^{2\ell(\alpha)}$ is the partition function
of the associated statistical mechanical model (Potts or O(n)).

We write the variational ansatz
\begin{equation}
|{\psi_k}\rangle = Z^{-1/2}{\sum_{\alpha}} e^{2\pi ik\,p({r_\alpha})} \:d^{\ell(\alpha)}\,|\alpha\rangle
\end{equation}
where $k$ is an integer.
These states are orthonormal because
\begin{equation}
\langle {\psi_k} | {\psi_l} \rangle = Z^{-1}{\sum_{\alpha}}
 d^{2\ell(\alpha)}\,e^{2\pi i p({r_\alpha}) (k-l)}
  = {\int_0^1} dp\,  e^{2\pi i p(k-l)}
  = \delta_{kl}
\end{equation}
In particular, $|{\psi_k}\rangle$ with $k\neq 0$ is orthogonal to
the ground state. Hence,
the expectation value of the Hamiltonian in this state is
an upper bound on the energy gap between the ground state
and the first excited state.
\begin{eqnarray}
\label{eqn:excited-state-energy1}
\Delta &\leq& \la \psi_k | H_d | \psi_k \ra - \la \psi_0 | H_d | \psi_0 \ra \cr &=&
\frac{1}{Z} {\sum_{\alpha,\beta}} 
\left(e^{2\pi i k (p({r_\alpha})-p({r_\beta}))} - 1 \right)\,d^{2\ell(\alpha)}
 \langle \alpha | H_d | \beta \rangle
 \end{eqnarray}
In the second line, we have used the fact that $\langle \alpha | H_d | \beta \rangle$
is non-zero only if $\ell(\alpha)=\ell(\beta)$.
Exchanging $\alpha$ with $\beta$ exchanges each term in
the sum in (\ref{eqn:excited-state-energy1}) with its complex conjugate.
Hence, we can write:
\begin{equation*}
\label{eqn:excited-state-energy2}
\Delta \leq 
2\,\text{Re}\left\{\frac{1}{Z} {\sum_{\alpha,\beta; {r_\beta}>{r_\alpha}}}\!\!\!
\left(e^{2\pi i k (p({r_\alpha})-p({r_\beta}))} - 1 \right)\,d^{2\ell(\alpha)}
 \langle \alpha | H_d | \beta \rangle\right\}
 \end{equation*}
Note that we have dropped the case ${r_\beta}={r_\alpha}$
since this expression vanishes for these configurations.
The $d$-isotopy Hamiltonian can change the length of a loop
by at most one lattice spacing, so $ \langle \alpha | H_d | \beta \rangle\neq 0$
only if ${r_\beta}={r_\alpha}+\frac{a}{L}$ where $a$ is the lattice spacing.
(Recall that $r_\alpha$ has been defined in units of $L$.)
Then, writing $p({r_\alpha})-p({r_\beta})\approx p'({r_\alpha})\,\frac{a}{L}$,
we have:
\begin{equation*}
\label{eqn:excited-state-energy3a}
\Delta \leq 
2\,\text{Re}\left\{\frac{1}{Z} {\sum_{\alpha,\beta; {r_\beta}>{r_\alpha}}}\!\!\!
\left(e^{2\pi i k p'({r_\alpha}){a}/{L}} - 1 \right)\,d^{2\ell(\alpha)}
 \langle \alpha | H_d | \beta \rangle\right\}
 \end{equation*}
Inserting ${\int_0^1} dr\, \delta(r-{r_\alpha})=1$ into this expression, we have:
\begin{eqnarray}
\label{eqn:excited-state-energy3b}
\Delta &\leq& 2\,\text{Re}\biggl\{
{\int_0^1} dr \,\delta(r-{r_\alpha})\frac{1}{Z} {\sum_{\alpha,\beta; {r_\beta}>{r_\alpha}}}\!\!\!
\left(e^{2\pi i k p'({r_\alpha}){a}/{L}} - 1 \right)\,\times \cr
& & {\hskip 4.5 cm} d^{2\ell(\alpha)}
 \langle \alpha | H_d | \beta \rangle\biggr\}\cr
 &=& 2\,\text{Re}\biggl\{
{\int_0^1} dr \,\left(1-e^{2\pi i k p'(r){a}/{L}}\right)\,{\overline \rho}(r)\biggr\}
\end{eqnarray}
where
\begin{eqnarray}
\label{eqn:f-def}
{\overline \rho}(r) &\equiv& -\frac{1}{Z} {\sum_{\alpha,\beta; {r_\beta}>{r_\alpha}}}\!\!\!
\delta(r-{r_\alpha})\,
d^{2\ell(\alpha)}\,
 \langle \alpha | H_d | \beta \rangle\cr
 &\equiv&
  \frac{1}{Z} {\sum_{\alpha}}
\delta(r-{r_\alpha})\,d^{2\ell(\alpha)}\,{\rho_\alpha}
 \end{eqnarray}
In the first line of (\ref{eqn:f-def}), we define ${\overline \rho}(r)$,
which is the expectation value of $\rho_\alpha$, defined in
the second line.
If we can argue that ${\overline \rho}(r)\leq A$ for some constant $A$
independent of $r$, then
\begin{eqnarray}
\label{eqn:excited-state-energy4}
\Delta &\leq& 2A\,\text{Re}\biggl\{
{\int_0^1} dr \,\left(1-e^{2\pi i k p'(r){a}/{L}}\right)\biggr\}\cr
&=& 2A \int dr \,\left(1-\cos\left({2\pi i k p'(r){a}/{L}}\right)\right)\cr
&\leq& 2A \left(1-\cos\left({2\pi i k M{a}/{L}}\right)\right)
\end{eqnarray}
Here, we have taken $p'(r)\leq M$, which follows from our
mild continuity assumptions.
Consequently, in the large $L$ limit, $\Delta \sim {k^2}/{L^2}$.
Note that the $L^{-2}$ reproduces the classical scaling for the
lowest eigenvalue of a string of length $L$, while the
$k^2$ dependence signals a quadratic dispersion relation associated
to a soft mode.

We are nearly finished. All that remains is to argue
that ${\overline \rho}(r)$ defined by (\ref{eqn:f-def})
satisfies ${\overline \rho}(r)\leq A$ for some constant $A$
independent of $r$. $\rho_\alpha$ is the
number of distinct sites (i.e. distinct terms in $H_d$) where $H_d$
can produce a fluctuation stretching out the ($x$-direction of) the
widest loop in the multi-loop $\alpha$ from breadth
$r_\alpha$ to breadth ${r_\alpha} + \frac{a}{L}$, i.e.
by one lattice step.
$\overline{\rho}(r)$ is the expectation value of
$\rho_\alpha$, averaged over all configurations
$\alpha$ whose widest loop has breadth $r$.
Note that in the scaling limit $\overline{\rho}(r)$
can have no $r$ dependence.  Prior to reaching
the scaling limit, we use argument $r$ in $\rho(r)$
to indicate maximum cluster breadth in units of $\frac{1}{L}$.
We now define $\rho_{\alpha,n}$ to be
$1$ if the widest loop $K\in\alpha$ meets the
right side of the unique smallest rectilinear box $B$,
containing $K$, $K \subset B$, in $n$
distinct ``fingers'' touching the right-hand wall
and zero otherwise.
We define $\overline{\rho_n}(r)$ to be its expectation
value,
\begin{equation}
\overline{\rho_n}(r) \equiv   \frac{1}{Z} {\sum_{\alpha}}
\delta(r-{r_\alpha})\,d^{2\ell(\alpha)}\,\rho_{\alpha,n}
\end{equation}
Then
\begin{equation}
\overline{\rho}(r) = {\sum_{n=1}^\infty} n\, \overline{\rho_n}(r)
\end{equation}

%%%%%%%%%%%%%%%%%%%%%%%%%%%%%%%%%%%%%%%%%%%%%%%%%%%%%%%%%%%
% \begin{figure}[hbt!]
% %\includegraphics[width=6.0cm]{fingers.eps}
% \caption{$K$ meets $B$ (at the right) in three fingers}
% \label{fig:fingers}
% \end{figure}
%%%%%%%%%%%%%%%%%%%%%%%%%%%%%%%%%%%%%%%%%%%%%%%%%%%%%%%%%%%%

The Hamiltonian $H_d$ can produce $n$ distinct states contributing to
${\overline{\rho}_1}({r +\frac{a}{L}})$ for each state contributing to
${\overline{\rho}_n}(r)$.  For
example, in the figure above, one may fluctuate any right-most
finger further to the right. Hence,
\begin{equation}
\label{eqn:rho-bound}
{\overline{\rho}_1}\!\left(r +\scriptstyle{\frac{a}{L}}\right) \geq  \sum_{n=1}^\infty n
{\rho_n}(r)
\end{equation}
But the right-hand-side is simply  $\overline{\rho}(r)$.
Since $\overline{\rho}^{}_{1}(r)$ is the ensemble average of
$\rho^{}_{\alpha,1}=0,1$, it must satisfy $0~\leq~\overline{\rho}^{}_{1}(r)~\leq~1$.
Hence, if there is a well-defined scaling limit, $\overline{\rho}^{}_{1}(r)\rightarrow\rho^{}_{1}$,
$\overline{\rho}(r)$ and
$\overline{\rho}({r +\frac{a}{L}})$ should converge. Then
the latter is the desired constant $A$
in the upper bound (\ref{eqn:excited-state-energy4})
on the excited state energy since we can re-write
(\ref{eqn:rho-bound}) as
\begin{equation}
\overline{\rho} \leq  \overline{\rho}^{}_{1}
\end{equation}

Naturally, this is only an upper bound, and in fact we have not
proved these are not just other degenerate ground states. However,
it seems unlikely that we have discovered ground state degeneracy
(on the sphere) since there is no continuous symmetry of the
Hamiltonian which could be broken or any other non-ergodicity.
Furthermore, we expect $\omega\propto k^2$ to be the correct
behavior of the low-lying excitations; our intuition is based on
observations made for other quasi-topological models
\cite{Henley97,Henley04,Ardonne04}, closely related to our $d=1$
case. Due to this characteristic quadratic spectrum, such lines of
critical points have been dubbed ``quantum Lifshitz
points''\cite{Ardonne04}.

Let us pause here and contemplate the physical reasons for having such
gapless modes. These excitations are \emph{not} Goldstone bosons,
since neither of the Hamiltonians
(\ref{eq:d-isotopy-Ham},\ref{eq:Mike-d-isotopy}) possesses any
continuous symmetry (which could be broken). Rather, these gapless
modes appear as a result of ``bottle-neck'' quantum dynamics which
only allows loops to fluctuate by either slowly growing or slowly
shrinking, one lattice plaquette at a time. As a result we can think
of configuration space as a very elongated, essentially
one-dimensional object parameterized by the diameter of the biggest
loop. While the quantum dynamics is in principle ergodic, in order to
reach a state with a long, order $L$ loop from a state with only short
loops, the entire ``length'' of this ``worm-like'' graph has to be
traversed.  From this analogy, we see that the eigenvalue problem for
our Hamiltonian is very similar to the eigenvalue problem for the
Laplacian operator on a string. (In contrast, geometries such as a
complete graph or hypercube do have spectral gaps; their bonds tie
them together more efficiently than links in a linear chain.)

Now, what about the spectrum when the Jones-Wenzl projectors are
implemented \cite{Freedman04a}? With the help of such additional
terms, going from short to long loops can be achieved by merging
existing loops together which can be done substantially faster than
by ``growing'' them.  These projectors would directly connect
various points of the ``worm-like'' configuration space of the
system.  In the case $d=1$ adding the two-strand projector
essentially reproduces Kitaev's ``toric code'' \cite{Kitaev97} and a
gap is opened. (We expect this to be the case for $d=-1$ as well.)
In the case of $n=d^2=2$, adding the three-strand projector is not
sufficient because the probability of three long loops coming
together within several lattice constants from each other -- a
necessary condition for a JW projector to act efficiently --
vanishes as $L^{-\alpha}$ for some $\alpha >
0$ \cite{Schramm-private}. Hence the effect of such terms on the
spectrum is too weak to open a gap (although it might lead to the
``stiffening'' of gapless excitations by reducing the dynamical
exponent $z$).

\section{Conclusions}
\label{sec:concl}

In this paper, we have analyzed the spectrum of low lying excitations
for a general class of local Hamiltonians whose ground state(s) are
characterized by $d$-isotopy. Using the statistical properties of their
non-local degrees of freedom we established the analog of a
Lieb-Schultz-Mattis theorem for quasi-topological systems. The
excitations are gapless, with the variational ansatz strongly
suggesting a quadratic ($\omega \propto k^2$) dispersion. What may
strike one as interesting and counter-intuitive is the fact that both
the Hamiltonian and the correlations of local operators are perfectly
short-ranged, however, the quantum dynamics is constrained to operate
on non-local, quasi-long-ranged objects -- loops -- which in turn
leads to a gapless spectrum. Finally, it is also interesting to remark
on the potential implications of such a behavior from the perspective
of understanding quantum glasses. The idea of using similar
quasi-topological models to model glassy behavior is not entirely new
\cite{Yin01,Yin02,Das01}, but until very recently \cite{Chamon05}, all
proposed models had a serious drawback, namely quasi-long range
correlations between \emph{local} degrees of freedom. This is contrary
to the fact that experimentally observed slow glassy dynamics has not
been accompanied by any divergent correlations. In this paper we have
explicitly demonstrated that such behavior is entirely possible in
the context of quasi-topological quantum critical points.

%%%%%%%%%%%%%%%%%%%%%%%%%%%%%%%%%%%%%%%%%%%%%%%%%%%%%%%%%%%%
%\begin{figure}[hbt]
%\includegraphics[width=2.5in]{}
%\caption{}
%\label{fig:}
%\end{figure}
%%%%%%%%%%%%%%%%%%%%%%%%%%%%%%%%%%%%%%%%%%%%%%%%%%%%%%%%%%%%

\begin{acknowledgments}
The authors would like to thank Oded Schramm for illuminating
discussions on the statistical properties of critical
configurations. We are also thankful to Matthew Hastings for pointing out
a gap in the proof presented in the earlier version of this manuscript. In
addition, we would like to acknowledge the hospitality of KITP and the
Aspen Center for Physics.
%\textbf{Anybody else?}
  C.~N.\ and K.~S.\ have been supported by the ARO under Grant
  No.~W911NF-04-1-0236.  C.~N.\ has also been supported by the NSF
  under Grant No.~DMR-0411800.
\end{acknowledgments}

%\bibliography{corr}
\bibliography{../bibs/corr,../bibs/reference}

\end{document}